\documentclass{llncs}

\usepackage[]{sosy-paper}
\usepackage{mdframed}
\usepackage[T1]{fontenc}
\usepackage{lmodern}

\usepackage[firstpage]{draftwatermark}
\SetWatermarkText{\hspace*{12.5cm}\raisebox{12.2cm}{\href{https://doi.org/10.5281/zenodo.5046084}{\includegraphics{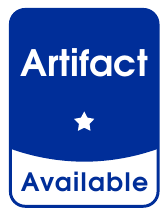}}}}
\SetWatermarkAngle{0}

\title{Towards a Benchmark Set for\\ Program Repair Based on Partial~Fixes}
\author{Dirk Beyer\inst{1}\orcidID{0000-0003-4832-7662},
        Lars Grunske\inst{2}\orcidID{0000-0002-8747-3745},
        Thomas Lemberger\inst{1}\orcidID{0000-0003-0291-815X}, and
        Minxing Tang\inst{2}\orcidID{0000-0003-0993-8874}}
\institute{LMU Munich, Germany \and
Humboldt-Universit\"at zu Berlin, Germany
}
\copyrightheader{}

\newcommand{\toolCol}{\texttt{col}\xspace}

\newcommand{\numRepositoriesConsidered}{\num{1500}\xspace{}}
\newcommand{\numRelevantRepositories}{\num{305}\xspace{}}
\newcommand{\numRelevantIssues}{\num{2204}\xspace{}}
\newcommand{\numPotentialIssues}{\num{2380}\xspace{}}
\newcommand{\numIssuesWithMissingBase}{\num{176}\xspace{}}

\colorlet{diffGreen}{green!20!white}
\colorlet{diffRed}{red!50!white}
\colorlet{diffMagenta}{violet!20!white}
\colorlet{diffGray}{lightgray!20!white}

\definecolor{delim}{RGB}{20,105,176}
\definecolor{numb}{RGB}{106, 109, 32}
\definecolor{string}{rgb}{0.64,0.08,0.08}

\newcommand\YAMLcolonstyle{\color{red}\mdseries}
\newcommand\YAMLkeystyle{\color{yellow}\bfseries}
\newcommand\YAMLvaluestyle{\color{blue}\mdseries}

\makeatletter

\newcommand\language@yaml{yaml}

\expandafter\expandafter\expandafter\lstdefinelanguage
\expandafter{\language@yaml}
{
  keywords={true,false,null,y,n},
  keywordstyle=\color{darkgray}\bfseries,
  basicstyle=\ttfamily\footnotesize\color{purple},
  numberstyle=\color{gray}\ttfamily\scriptsize,
  sensitive=false,
  comment=[l]{\#},
  morecomment=[s]{/*}{*/},
  commentstyle=\color{darkgray}\ttfamily,
  stringstyle=\YAMLvaluestyle\ttfamily,
  moredelim=[l][\color{orange}]{\&},
  moredelim=[l][\color{magenta}]{*},
  moredelim=**[il][\YAMLcolonstyle{:}\YAMLvaluestyle]{:},   
  morestring=[b]',
  morestring=[b]",
  literate =    {---}{{\ProcessThreeDashes}}3
                {>}{{\textcolor{red}\textgreater}}1
                {|}{{\textcolor{red}\textbar}}1
                {\ -\ }{{\mdseries\ -\ }}3,
}

\lst@AddToHook{EveryLine}{\ifx\lst@language\language@yaml\YAMLkeystyle\fi}
\makeatother

\newcommand\ProcessThreeDashes{\llap{\color{cyan}\mdseries-{-}-}}

\lstdefinelanguage{json}{
    numbers=left,
    numberstyle=\tiny,
    stringstyle=\scriptsize\color{darkgray},
    rulecolor=\color{black},
    showspaces=false,
    showtabs=false,
    breaklines=true,
    postbreak=\raisebox{0ex}[0ex][0ex]{\ensuremath{\color{gray}\hookrightarrow\space}},
    breakatwhitespace=true,
    basicstyle=\ttfamily\scriptsize,
    upquote=true,
    morestring=[b]",
    literate=
     *{0}{{{\color{numb}0}}}{1}
      {1}{{{\color{numb}1}}}{1}
      {2}{{{\color{numb}2}}}{1}
      {3}{{{\color{numb}3}}}{1}
      {4}{{{\color{numb}4}}}{1}
      {5}{{{\color{numb}5}}}{1}
      {6}{{{\color{numb}6}}}{1}
      {7}{{{\color{numb}7}}}{1}
      {8}{{{\color{numb}8}}}{1}
      {9}{{{\color{numb}9}}}{1}
      {\{}{{{\color{delim}{\{}}}}{1}
      {\}}{{{\color{delim}{\}}}}}{1}
      {[}{{{\color{delim}{[}}}}{1}
      {]}{{{\color{delim}{]}}}}{1},
}

\begin{document}

\maketitle

\begin{abstract}
Software bugs significantly contribute to software cost
and increase the risk of system malfunctioning.
In recent years, many automated program-repair approaches have been proposed to automatically fix undesired program behavior. Despite of their great success, specific problems such as fixing bugs with partial fixes still remain unresolved.
A~partial fix to a known software~issue is a programmer's failed attempt to fix the issue the first time.
Even though it fails, this fix attempt still conveys important information such as the suspicious software region and the bug type.
In this work we do not propose an approach for program repair with partial fixes,
but instead answer a preliminary question: Do partial fixes occur often enough, in general,
to be relevant for the research area of automated program repair?
We crawled \num{1500}~open-source C~repositories on GitHub for partial fixes.
The result is a benchmark set of \numRelevantIssues~benchmark tasks for automated program repair
based on partial fixes.
The benchmark set is available open source and open to further contributions and improvement.

\end{abstract}


\section{Introduction}

A vast amount of software-development resources are claimed
by debugging (locating and fixing software bugs).
To make developers more productive in the process,
different techniques exist:
Software testing~\cite{ArtOfSoftwareTesting} and formal verification~\cite{HBMC-book}
figure out whether bugs exist \emph{somewhere},
and automated fault localization~\cite{FaultLocalizationSurvey} can propose
code locations that may be buggy.
But even fixing a known bug seems to pose challenges:
Multiple studies~\cite{SupplementaryFixes,FirefoxPartialFixes,AndroidPartialRepair,EclipseMozillaPartialFixes,PartialFixesJavaManualInspection,
PartialFixesOperatingSystems,PartialFixesInSmallChanges}
on selected 
software projects showed
a large number of fixes 
in these projects
did not actually fix the bug.
(Such fixes are also known as partial or incomplete~patches.)

Automated program repair~\cite{ProgramRepairSurvey}
repairs bugs in programs, so it may also
help with the task of identifying these partial fixes
and providing one or more supplementary patches%
.
But automated program repair is still ongoing research.
Confronting a large number of partial fixes in software
and the existing challenges of automated program repair,
it is worthwhile to consider a specialized type of automated program repair,
aimed at partial fixes:
If a software developer tries to fix a bug and fails,
this partial fix conveys valuable information about
the relevant code location and the program semantics the developer
suspects to be wrong.
This information may be exploited by specialized techniques
of automated program repair to generate better patches. 

Multiple benchmark sets exist for automated program repair~\cite{Defects4J, iBugs, CoreBench}
and the existence of partial fixes has already been studied~\cite{FirefoxPartialFixes,AndroidPartialRepair,EclipseMozillaPartialFixes,PartialFixesJavaManualInspection,
PartialFixesOperatingSystems}.
Unfortunately, a benchmark set
for evaluating automated program repair with partial fixes does not exist. This
makes it impossible to evaluate new approaches
that aim to leverage partial fixes.
%
In addition, no work has yet explored the prevalence of partial fixes in C
on a large number of code repositories;
only hand-selected, large projects were considered in previous approaches.

We were able to collect year-long experience
by maintaining the sv-benchmarks\footnote{\url{https://github.com/sosy-lab/sv-benchmarks/}}
benchmark~set, the largest available benchmark set for automated software verification of C~programs.
We use this experience to propose the first benchmark~set for automated program repair
for partial fixes in C.
To obtain a large number of benchmark tasks, we examined the \num{1500}~most-starred GitHub repositories
and applied two selection heuristics to extract partial fixes.
This yields a set of \numRelevantIssues~candidate benchmark tasks for further research
in the area of automated program repair with partial fixes.

\subsection*{Example}

\begin{figure}[t]
    \begin{mdframed}[backgroundcolor=diffGray, linewidth=0, roundcorner=10pt, leftmargin=0pt, rightmargin=0pt]
    \begin{lstlisting}[language=C,numbers=none,xleftmargin=0pt,basicstyle=\footnotesize\ttfamily]
(*@\colorbox{diffMagenta}{@@@ -391,6 +391,8 @@ int main(int argc, char **argv)~~~~~~~~~~~~~~~~~} @*)
/* goto the last line that had a character on it */
for (; l->l_next; l = l->l_next)
  this_line++;
(*@\colorbox{diffGreen}{+~if (max_line == 0)~~~~~~~~~~~~~~~~~~~~~~~~~~~~~~~~~~~~~~~~~~~~~~~}@*)
(*@\colorbox{diffGreen}{+~~~return EXIT_SUCCESS;~~~~/* no lines, so just exit */~~~~~~~~~~~}@*)
flush_lines(this_line - nflushd_lines + extra_lines + 1);
    \end{lstlisting}
    \end{mdframed}
    \vspace{-3mm}

    \caption{First, partial fix to issue~422 with \toolCol}
    \label{fig:colFix1}
    \vspace{5mm}

    \begin{mdframed}[backgroundcolor=diffGray, linewidth=0, roundcorner=10pt, leftmargin=1pt, rightmargin=1pt]
    \begin{lstlisting}[language=C,numbers=none,xleftmargin=0pt,basicstyle=\footnotesize\ttfamily]
(*@\colorbox{diffMagenta}{@@ -396,7 +396,7 @@ int main(int argc, char **argv)~~~~~~~~~~~~~~~~~~} @*)
/* goto the last line that had a character on it */
for (; l->l_next; l = l->l_next)
  this_line++;
(*@\colorbox{diffRed}{-~if (max_line == 0)~~~~~~~~~~~~~~~~~~~~~~~~~~~~~~~~~~~~~~~~~~~~~~~}@*)
(*@\colorbox{diffGreen}{+~if (max_line == 0 \&\& cur_col == 0)~~~~~~~~~~~~~~~~~~~~~~~~~~~~~~~}@*)
  return EXIT_SUCCESS;    /* no lines, so just exit */
flush_lines(this_line - nflushd_lines + extra_lines + 1);
    \end{lstlisting}
    \end{mdframed}
    \vspace{-3mm}

    \caption{Second, successful fix to issue~422 with \toolCol}
    \label{fig:colFix2}
\end{figure}

GitHub repository \href{https://github.com/karelzak/util-linux}{karelzak/util-linux}
consists of different command-line tools for GNU/Linux.
One of these tools is \toolCol,
which removes the unicode characters `reverse line feed' (go up one line)
and `half-reverse line feed' (go up half a line) from a given input.

In revision \href{https://github.com/karelzak/util-linux/commit/c6b0cbdd95ff30788cdb6afee707e20f0dd640b8}{c6b0cb},
\toolCol falsely printed a newline if the input was~empty.
Issue~422\,\footnote{\url{https://github.com/karelzak/util-linux/issues/422}}
manages this bug.
As shown in \cref{fig:colFix1}, a first attempt to fix the bug failed.
This attempt added new code that tries to implement the expected behavior,
but the checked variable \lstinline{max_line} only counts lines with a line break at their end,
so a single line of program~input without a line break leads
the new revision to exit before printing output.
Later, a second attempt fixed this new issue (\cref{fig:colFix2})
by adjusting the added check accordingly.

Software bugs like our example may be difficult
to fix by automated program repair: Without the partial fix,
it is difficult to identify both the expected behavior
(\lstinline{return EXIT_SUCCESS})
and the ``best'' location for adding
the additional code.
With both information inferred by the partial fix,
automated program repair approaches can focus
on improving the affected program behavior.

\section{Identification of Partial Fixes}

\subsection{Repository Selection}

We select repositories based on star count.
We consider the \numRepositoriesConsidered~most-starred C~repositories
on \href{https://github.com}{GitHub}.

\subsection{Issue Selection}

For each repository, we crawl all issues.
We only consider closed issues with at least two associated commits.
We need two commits as one commit can be the failed attempt and the other one can be the final fix.
A commit is associated with a certain issue if it refers to the issue number in its commit message
or if its commit hash is mentioned in the comment of this issue.
For each issue, we identify partial fixes based on two individual patterns:
\emph{reopen-close} and \emph{fail-fix}.

\paragraph{Reopen-Close Pattern}

We consider an issue to contain a partial~fix if:
(1)~there is a commit associated with the issue,
(2)~the issue is closed afterwards,
(3)~the issue is then reopened,
(4)~a second commit occurs, and
(5)~the issue is closed at the end.
Our introductory example fulfills this pattern.
To
avoid accidental closes, we only consider issues that were reopened
at least 5~minutes after the first close.
If the reopen-close pattern occurs multiple times in an issue,
we only consider the last reopen and the last close.

\paragraph{Fail-Fix Pattern}

In addition, we consider an issue to contain a partial~fix if:
(1)~there is a commit associated with the issue, and the continuous-integration~(CI) tests fail for this commit,
(2)~there is a later commit associated with the issue, and the CI tests succeed for this commit, and
(3)~the issue is closed at the end.

\bigskip
Of the \numRepositoriesConsidered~repositories considered,
\numRelevantRepositories~contain at least one issue that matches at least one of our patterns,
and \numPotentialIssues~issues match at least one of our patterns.
Note that the remaining \num{1195}~repositories without any match may contain
partial fixes that we can not identify:
We can only identify issues that are connected by references.
A reference is implied by (a)~a comment in the issue that mentions the commit explicitly,
or (b)~a reference to the issue in the commit message (e.g., `\textit{fixes \#1}').
That means that we miss all repositories with a workflow that does not put commits and issues
in relation. For example, if a developer does fix the issue, but closes the issue manually,
there is no direct relation between the fix and the issue, so it is not discovered by our
patterns.
We also do not consider merge requests, which are often used to fix issues. In these workflows,
the commits themselves do not reference the issue, but only the merge-request description does.
%

\subsection{Issue Metadata}

\begin{figure}[t!]
    \lstset{language=json}
    \begin{lstlisting}
{
  "commits": [
    {
      /* ... snip ... */
      "is_parent": true,
      "sha": "c6b0cbdd95ff30788cdb6afee707e20f0dd640b8"
    },
    {
      "diff": "karelzak_util-linux/issue_422/commit_b6b5272.diff",
      "diff_additions": 2,
      "diff_deletions": 0,
      "diff_file_size": 18794,
      "exists": true,
      "files": [
        {
          "file_additions": 2,
          "file_deletions": 0,
          "file_exists": true,
          "file_name": "text-utils/col.c",
          "file_size": 18794
        }
      ],
      "interesting": {
        "reopen": true,
        "status": false
      },
      "is_parent": false,
      "sha": "b6b5272b03ea9d3fa15601801d4d0f76ea4440f1"
    },
    /* ... snip more commits ... */
  ],
  "first_commit": "2017-05-10 08:53:28",
  "last_commit": "2020-06-20 21:17:30",
  "number": 422,
  "score": {
    "all_labels": [],
    "fix_size": 2,
    "intersecting": {
      "computed_score": 1.0,
      "final_commit_size": 2,
      "intersection_size": 2,
      "previous_commits_size": 4
    },
    "is_makefile_based": true,
    "total_changes": 8
  }
}
    \end{lstlisting}
    \caption{Excerpt of collected issue metadata}
    \label{fig:jsonMetadata}
\end{figure}

For each of the \numPotentialIssues~issues
that contains a partial fix, we collect the following metadata
and store it as individual JSON~file (cf.~\cref{fig:jsonMetadata}):
     All commits related to the issue,
     date and time of the first commit,
     date and time of the last commit,
     labels associated with the issue, and
    the total number of lines changed by the commits associated with the issue.

\section{Partial-Fix Benchmark Set}


\subsection{Adopted Best Practices}

\newcommand{\smallsec}[1]{\vspace{2mm}\noindent\textit{#1}.}

From our experience with sv-benchmarks%
\,\footnote{\url{https://github.com/sosy-lab/sv-benchmarks/}}%
, we adopt the following practices:

\smallsec{Unique Task Names}
Each benchmark task should be uniquely identifiable from its name.
This eases communication about tasks and task handling.

\smallsec{YAML Task-Definition Format}
We use separate files for the definition of a benchmark task
and the program files used for program repair.
This makes maintenance of task definitions
and the future addition of metadata
and additional features easy.
We chose YAML as language for our task-definition format
as it is easily readable and widely supported.

\smallsec{Open Source}
A benchmark set requires constant maintenance and strives
from community contributions.
To that means, we publish all tools and data surrounding the benchmark set
as well as the benchmark set itself under the permissive Apache license~2.0 and on \href{https://gitlab.com}{GitLab}.

\subsection{Benchmark Set}

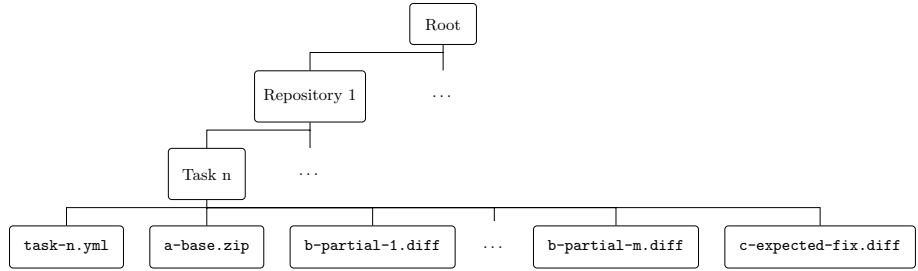
\begin{figure}[t]

    \resizebox{\linewidth}{!}{
    \begin{tikzpicture}[
      node distance=5mm,
        every node/.style={rounded corners=2pt,},
        item/.style={rectangle, inner sep=5pt, minimum height=.8cm, minimum width=1.3cm},
        repo/.style={item, minimum width=2cm, minimum height=1cm},
        issue/.style={item, minimum width=1.5cm, minimum height=1cm},
        file/.style={draw, inner sep=8pt, font=\ttfamily},
        ]
        \node[item, draw] (root) {Root};
        \node[repo, below = of root] (repodots) {\ldots};
        \node[draw,repo, left = of repodots] (repo1) {Repository 1};

        \node[issue, below = of repo1] (issuedots) {\ldots};
        \node[draw, issue, left = of issuedots] (issue1) {Task n};

        \node[file, below = of issue1] (original) {a-base.zip};
        \node[file, left = of original] (yml) {task-n.yml};
        \node[file, right = of original] (attempt1) {b-partial-1.diff};
        \node[issue, right = 0mm of attempt1] (attemptDot) {\ldots};
        \node[file, right = 0mm of attemptDot] (attemptN) {b-partial-m.diff};
        \node[file, right = of attemptN] (expected) {c-expected-fix.diff};

        \coordinate (firstmiddle) at ($(root.south)!.3!(repodots.north)$);
        \coordinate (secondmiddle) at ($(repo1.south)!.3!(issuedots.north)$);
        \coordinate (thirdmiddle) at ($(issue1.south)!.3!(original.north)$);

        \path[draw] (root) -- (firstmiddle) -| (repo1)
                    (firstmiddle) -- (repodots)
                    (repo1) -- (secondmiddle) -| (issue1)
                    (secondmiddle) -| (issuedots)
                    (issue1) -- (thirdmiddle) -| (yml)
                    (thirdmiddle) -| (original)
                    (thirdmiddle) -| (attempt1)
                    (thirdmiddle) -| (expected)
                    (thirdmiddle) -| (attemptDot)
                    (thirdmiddle) -| (attemptN);
    \end{tikzpicture}
    }
    \caption{Directory and file structure of the created partial-fix benchmark set}
    \label{fig:repoStructure}
\end{figure}

Using the collected data,
we download all relevant program versions and changes.
If a program version is not available
(e.g., because the change was done in some unavailable fork),
we skip that issue.
We skip \numIssuesWithMissingBase~issues, so our
final benchmark set consists of \numRelevantIssues~benchmark tasks.

The benchmark set has the following structure (\cref{fig:repoStructure}):
All tasks that are created from the same repository are grouped 
in a directory that is named by the repository owner and the repository name.
For example, for owner karelzak and repository util-linux,
the directory name is \texttt{karelzak_util-linux}.
Each task is in an own directory that is named according to the issue number,
e.g., \texttt{issue_422}.
In that task directory, a uniquely named YAML file (e.g., \texttt{karelzak_util-linux_422.yml}) defines the benchmark task.
The remaining files are
the repository state before the attempted fix (\texttt{a-base.zip}),
the changes introduced by the attempted fix (\texttt{b-partial-1.diff} to \texttt{b-partial-n.diff}),
and the final fix (\texttt{c-expected-fix.diff}).
Note that there may be multiple partial fixes before the final fix was introduced.
For each separation of partial fixes and final fix a new benchmark task could be created,
by merging the last $n$ partial fixes into the expected fix.
For example, 
if there are two partial fixes \texttt{b-partial-1.diff}
and \texttt{b-partial-2.diff},
changes \texttt{b-partial-1.diff} and \texttt{b-partial-2.diff}
could be considered as the partial fix and \texttt{c-expected-fix.diff}
could be considered the final fix.
But another task could consider only \texttt{b-partial-1.diff}
the partial fix and consider changes \texttt{b-partial-2.diff} and \texttt{c-expected-fix.diff}
the final fix.
In this work, we always only use the last change to the program (\texttt{c-expected-fix.diff})
as final fix, to keep complexity low.

\subsection{Task-Definition Format}

\begin{figure}[t]
    \begin{lstlisting}[language=yaml]
# Version of this task-definition format - will be increased
# with future changes.
format_version: '1.0'

# URL of the base repository with the bug
repository_url: https://github.com/karelzak/util-linux
# URL to the issue that describes the related bug
issue_url: https://github.com/karelzak/util-linux/issues/422


# Information supposed to be given to the repair tool
input_files:
    # The base version of the program-under-repair, with
    # the original bug.
    base_version:
        # archive of program
        input_file: 'a-base.zip'
        # commit hash of the base version 
        commit-sha1: c6b0cbd
    # The first fix attempt. This is a list of
    # one or more changes that tried to fix the bug.
    fix_attempt:
        - input_file: 'b-partial-1.diff'
          commit-sha1: b6b5272

# The expected fix.
expected_fix:
    input_file: 'c-expected-fix.diff'
    commit-sha1: d8bfcb4

# Metadata about the task. At the moment, only 'language',
# to specify the code language of the benchmark task.
options:
    language: C
    \end{lstlisting}
\caption{Example task-definition file}
\label{fig:yaml}
\end{figure}

\Cref{fig:yaml} shows an example task definition for our introductory example
with explanations of each field.
For benchmarking, the input file of the \texttt{base_version}
and the input files of \texttt{fix_attempt}
are given as input to a program-repair tool.
The base version is an archive of the full checkout of the program in the
buggy version.
The \texttt{fix_attempt} is a list of changes that attempted to fix
the bug, but failed or introduced new bugs.
Each change is represented as a diff file that represents the changes
as line~additions and deletions (as in \cref{fig:colFix1}).
When the program-repair tool proposes a fix,
the benchmarking infrastructure can compare this fix against
the \texttt{expected_fix} (also a diff file) to check for correctness.

\section{Future Work}

\paragraph{Oracle Creation}
Automated program repair requires an oracle to distinguish
desired from undesired program behaviors.
Program specifications and
executable test suites are examples for oracles.
Our benchmark set does not yet contain oracles for the benchmark tasks.
Automated oracle creation is difficult and an active research topic.

\paragraph{Validation}
We have not yet validated our benchmark set.
To make sure that the created benchmark tasks are meaningful,
we will run existing tools for automated program repair,
for example, Angelix~\cite{Angelix} or CPR~\cite{CPR}, on our benchmark set.

\paragraph{Improving Categorization}
We try to provide a categorization of benchmark tasks for a better overview
of the benchmark tasks.
Unfortunately, the categorization is time-consuming
and attempted machine-learning approaches are not yet trustful enough.
This could be improved to provide a more confident categorization of benchmark tasks.

\paragraph{Adding Benchmark Tasks}
To obtain more benchmark tasks, we should consider multiple options:
(1)~Merge requests in addition to issues. Often, merge requests are directly used
to fix issues that are either not reported, or the commits of the merge request do not explicitly
mention the issue they target.
(2)~Further patterns.
We identified the following additional pattern for partial fixes,
that we do not consider yet:
\textit{A developer creates multiple commits on a branch or fork, all targeting at an issue.
But only a subset of these commits are merged into the master for closing this issue.}
Either all but the last commit could be considered as partial fixes,
or only the commits not ending up in the final merge could be the partial fixes.

\section{Conclusion}

We have presented the first large benchmark set for automated program repair
with partial fixes in C programs,
with \numRelevantIssues~benchmark tasks.
This creates a baseline for future research in this new, promising research field.

\subsection*{Data Availability Statement}

Our benchmark set is open-source and maintained at \href{https://gitlab.com/sosy-lab/research/data/partial-fix-benchmarks}{https://gitlab.com/sosy-lab/research/data/partial-fix-benchmarks}.
The software used to collect and create the benchmark set is open-source
and maintained at \href{https://gitlab.com/sosy-lab/software/partial-fix-benchmarks/}{https://gitlab.com/sosy-lab/software/partial-fix-benchmarks/}.
The version used in this work is \href{https://gitlab.com/sosy-lab/research/data/partial-fix-benchmarks/-/tree/arXiv-v1}{\texttt{arXiv-v1}}.
It is archived and available at Zenodo\,\cite{PartialFixBenchmarkSet-artifact}.

\bibliography{dbeyer,sw,artifact}

\providecommand{\serysort}{}
\begin{thebibliography}{10}
\providecommand{\url}[1]{\texttt{#1}}
\providecommand{\urlprefix}{URL }
\providecommand{\doi}[1]{https://doi.org/#1}

\bibitem{SupplementaryFixes}
An, L., Khomh, F., Adams, B.: Supplementary bug fixes vs. re-opened bugs. In:
  Proc.\ {SCAM}. pp. 205--214. IEEE (2014). \doi{10.1109/SCAM.2014.29}

\bibitem{PartialFixBenchmarkSet-artifact}
Beyer, D., Grunske, L., Lemberger, T., Tang, M.: {Partial-fix benchmarks}: A
  benchmark set for program repair on partial fixes (2021).
  \doi{10.5281/zenodo.5046084}

\bibitem{CoreBench}
B{\"{o}}hme, M., Roychoudhury, A.: {CoREBench}: {S}tudying complexity of
  regression errors. In: Proc.\ {ISSTA}. pp. 105--115. {ACM} (2014).
  \doi{10.1145/2610384.2628058}

\bibitem{HBMC-book}
Clarke, E.M., Henzinger, T.A., Veith, H., Bloem, R.: Handbook of Model
  Checking. Springer (2018). \doi{10.1007/978-3-319-10575-8}

\bibitem{iBugs}
Dallmeier, V., Zimmermann, T.: Extraction of bug-localization benchmarks from
  history. In: Proc.\ {ASE}. pp. 433--436. {ACM} (2007).
  \doi{10.1145/1321631.1321702}

\bibitem{ProgramRepairSurvey}
Gazzola, L., Micucci, D., Mariani, L.: Automatic software repair: {A} survey.
  {IEEE} Trans. Software Eng.  \textbf{45}(1),  34--67 (2019).
  \doi{10.1109/TSE.2017.2755013}

\bibitem{AndroidPartialRepair}
Gu, Z., Barr, E.T., Hamilton, D.J., Su, Z.: Has the bug really been fixed? In:
  Proc.\ {ICSE}. pp. 55--64. ACM (2010). \doi{10.1145/1806799.1806812}

\bibitem{Defects4J}
Just, R., Jalali, D., Ernst, M.D.: {Defects4J}: {A} database of existing faults
  to enable controlled testing studies for {Java} programs. In: Proc.\ {ISSTA}.
  pp. 437--440. {ACM} (2014). \doi{10.1145/2610384.2628055}

\bibitem{Angelix}
Mechtaev, S., Yi, J., Roychoudhury, A.: Angelix: {S}calable multiline program
  patch synthesis via symbolic analysis. In: Proc.~{ICSE}. pp. 691--701. {ACM}
  (2016). \doi{10.1145/2884781.2884807}

\bibitem{ArtOfSoftwareTesting}
Myers, G.J., Sandler, C., Badgett, T.: The Art of Software Testing. Wiley, 3rd
  edn. (2011)

\bibitem{PartialFixesJavaManualInspection}
Nguyen, T.T., Nguyen, H.A., Pham, N.H., Al{-}Kofahi, J.M., Nguyen, T.N.:
  Recurring bug fixes in object-oriented programs. In: Proc.\ {ICSE}. pp.
  315--324. {ACM} (2010). \doi{10.1145/1806799.1806847}

\bibitem{EclipseMozillaPartialFixes}
Park, J., Kim, M., Ray, B., Bae, D.: An empirical study of supplementary bug
  fixes. In: Proc.\ {MSR}. pp. 40--49. IEEE (2012).
  \doi{10.1109/MSR.2012.6224298}

\bibitem{PartialFixesInSmallChanges}
Purushothaman, R., Perry, D.E.: Toward understanding the rhetoric of small
  source-code changes. {IEEE} Trans. Software Eng.  \textbf{31}(6),  511--526
  (2005). \doi{10.1109/TSE.2005.74}

\bibitem{FirefoxPartialFixes}
Seo, H., Kim, S.: Predicting recurring crash stacks. In: Proc.\ {ASE}. pp.
  180--189. {ACM} (2012). \doi{10.1145/2351676.2351702}

\bibitem{CPR}
Shariffdeen, R.S., Noller, Y., Grunske, L., Roychoudhury, A.: Concolic program
  repair. In: Proc. PLDI. pp. 390--405. {ACM} (2021).
  \doi{10.1145/3453483.3454051}

\bibitem{FaultLocalizationSurvey}
Wong, W.E., Gao, R., Li, Y., Abreu, R., Wotawa, F.: A survey on software fault
  localization. {IEEE} Trans. Software Eng.  \textbf{42}(8),  707--740 (2016).
  \doi{10.1109/TSE.2016.2521368}

\bibitem{PartialFixesOperatingSystems}
Yin, Z., Yuan, D., Zhou, Y., Pasupathy, S., Bairavasundaram, L.N.: How do fixes
  become bugs? In: Proc.\ {ESEC/FSE}. pp. 26--36. {ACM} (2011).
  \doi{10.1145/2025113.2025121}

\end{thebibliography}

\end{document}